\documentclass[a4paper,11pt]{scrartcl}

\usepackage[margin=2cm]{geometry}
\usepackage{booktabs}
\usepackage{multirow}
\usepackage{amssymb}
\usepackage{rotating}
\usepackage{multicol}
\usepackage{tabularx}
\usepackage{authblk}
\usepackage{footnote}
\usepackage[hidelinks]{hyperref}

\newcolumntype{C}[1]{>{\centering\let\newline\\\arraybackslash\hspace{0pt}}m{#1}}
\newcolumntype{R}[1]{>{\raggedleft\let\newline\\\arraybackslash\hspace{0pt}}m{#1}}

\newcommand{\matchertitle}[1]{\vspace{1em} \noindent \textbf{#1.}}

\usepackage{natbib}
\begin{document}

\title{How Process Model Matching Techniques\\ Use Control Flow Information} 
\subtitle{Online Appendix to:\\Analyzing Control Flow Information to Improve the Effectiveness of Process Model Matching Techniques}

\author[1,2]{Christopher Klinkm\"uller}
\author[2,3]{Ingo Weber}
\affil[1]{Department of Computing, Macquarie University, Sydney, Australia}
\affil[2]{Data61, CSIRO, Sydney, Australia}
\affil[3]{University of New South Wales, Sydney, Australia}

\date{}

\maketitle

\section{Introduction}

Process model matchers automate the identification of correspondences between process models, i.e., activities that represent similar functionality in different models.
This way they support a variety of tasks in \textit{business process management}, e.g., the management of process model collections, the consolidation of processes, or the re-use of process fragments at design time.
To detect correspondences, matchers generally compare activity labels which provide brief natural language descriptions.
Additionally, some matchers attach importance to the control flow in the process models and analyze the temporal dependencies between the activities.
Yet, the process model matching contests in 2013 and 2015 \citep{Cayoglu+.2014,Antunes+.2015} revealed that the effectiveness of state-of-the-art matchers is low, i.e., their results only contain a few existing and many irrelevant correspondences.

With that in mind, we aim to provide guidance for the development of more effective matchers.
In particular, we here examine the question: \textit{how can process model matchers benefit from the analysis of control flow information?}
To answer this question, we review the process model matching literature in order to identify options for using control flow information in the matching process.
As we are interested in improving the matcher effectiveness, we further assess the evidence that was given towards the successful contribution of control flow information to the matching process.
We limit our focus to control flow information, because the comparison of labels has been extensively studied in natural language processing \citep{ManningSchutze.1999}, information retrieval \citep{Manning+.2008}, as well as schema and ontology matching \citep{EuzenatShvaiko.2007,RahmBernstein.2001}.

The remainder of this report is organized as follows.
We first outline how we identified process model matching literature in Section~\ref{sec:search}.
Next, we provide a brief overview of the identified literature in Section~\ref{sec:overview}.
Then, we present a classification of analysis options that we derived from the literature and discuss the evidence given towards them in Section~\ref{sec:results}.
Finally, Section~\ref{sec:conclusion} concludes the report.
\section{Identification of Literature}
\label{sec:search}

In this section, we outline the literature search process that was applied to identify relevent publications.
This search process was carried out by the first author in the context of his PhD thesis \citep{Klinkmuller.2017} where he reviewed the process model matching literature with regard to the general applicability of matchers, their effectiveness, and the applied research approaches.
In this work, we extend this review by analyzing the use of control flow information in those matchers.

To identify relevant publications,
guidelines for literature reviews \citep{vomBrocke+.2009} were applied. 
%
In the first step, the matching contests \citep{Cayoglu+.2014,Antunes+.2015} which were considered as a source of relevant works were analyzed.
Here, six additional publications that introduced process model matching techniques were identified and an initial set of eight publications was obtained.
This set served as a basis for the verification of the search strategy in a later step.
Then, the database search was prepared.
As the overall goal was the identification of papers that introduce matching techniques, ``process model matching'' was chosen as the basic search string.
Yet, this term is quite specific and might not cover all relevant publications.
Hence, it was split into the terms ``process model'' and ``matching'' for each of which synonyms were considered.
That is, works that deal with process models or workflows, and refer to matching, mapping, or alignment were considered, too.
Therefore, the combined search string is [(``process model'' OR ``workflow'') AND (``match*'' OR ``align*'' OR ``map*'')]. 
Moreover, the search was restricted to scientific, peer-reviewed, English literature that was published between January 2000 and May 2016,
where the former date is associated with the rise of modern BPM and the latter marks the time at which the review was completed. 

Next, the database search was carried out which relied on the following databases that comprise publications from the information systems domain: Springer Link, ACM Digital, IEEE Xplore, Science Direct, Emerald Insight, and Google Scholar.\footnote{Each database uses a proprietary format for search criteria, to which we adapted the format of the search string and cut-off dates.}
Querying the databases resulted in lists that comprised hundreds or thousands of papers, where papers with a low rank were highly likely to be irrelevant.
Given the specific scope of the search, each list was scanned stepwise starting from the highest rank, and as per common practice stopped scanning when the distance to the last relevant paper exceeded 20. At a minimum, the first 50 entries were considered.
In total, 2261 entries were reviewed and 47 papers were marked as relevant based on a title and abstract scan.

At this point, the search strategy was evaluated by checking if publications known to be relevant were included. 
For this check, the eight initially identified works were considered.
When checking whether this eight papers were part of the 48 results, it was found that,
except for \citep{Antunes+.2015}, all publications were included.
This paper 
was not found since it was published in the Lecture Notes in Informatics 
through the German society for computer science (Gesellschaft f{\"u}r Informatik) which were not indexed by any of the databases at this time.
This result was considered to verify the strategy, but its completeness is still limited by the completeness of the employed databases.
To mitigate the risk of overlooking papers, the search was complemented by a backward search over the references in the identified papers.
Here, two more relevant publications were identified.

Finally, the identified papers were filtered and only those that were relevant with respect to the scope were selected.
For this step, the inclusion criterion was that the papers introduced a process model matcher.
On the contrary, papers were excluded if they (i) discussed process model matching; (ii) addressed aspects of model collection management; 
(iii) discussed support for process model design; or (iv) referred to Business-IT alignment.
Additionally, the publications \citep{Gerth+.2011,Gerth.2014} were identified as duplicates.
The remaining 19 papers all propose a matcher.
Tables~\ref{tab:overviewI} and \ref{tab:overviewII} summarize the classification and source of all 49 publications that were considered during the search process.

\begin{table}[hb]
\caption{Publications considered during the literature (part I)}
\label{tab:overviewI}
\centering
\begin{tabular}{l || c || c }
\hline
Reference & Topic & First Source\\
\hline
\hline
\citep{Zhuge.2002}            & Collection Management   & Springer \\
\hline
\citep{Wombacher+.2003}       & Collection Management   & IEEE Explore \\
\hline
\citep{Wombacher+.2004}       & Collection Management   & IEEE Explore \\
\hline
\citep{Brockmans+.2006}       & Model Matching          & \citep{Dijkman+.2009}\\
\hline
\citep{SuwannopasSenivongse.2006} & Collection Management   & Google Scholar\\
\hline
\citep{Lei+.2007}             & Collection Management   & Springer\\
\hline
\citep{Nejati+.2007}          & Model Matching          & \citep{Dijkman+.2009}\\
\hline
\citep{DeutchMilo.2009}       & Collection Management   & IEEE Explore \\
\hline
\citep{Dijkman+.2009}         & Model Matching          & Matching Contest \\
\hline
\citep{GacituaDecarPahl.2009} & Collection Management		& IEEE Explore \\
\hline
\citep{GaoZhang.2009}         & Collection Management   & ACM Digital \\
\hline
\citep{Jung.2009}             & Collection Management   & ACM Digital \\
\hline
\citep{ZhuPung.2009}					& Collection Management   & IEEE Explore \\
\hline
\citep{AkkirajuIvan.2010}		  & Collection Management		& Springer \\
\hline
\citep{GacituaDecarPahl.2010} & Collection Management   & IEEE Explore \\
\hline
\citep{Gater+.2010b}          & Collection Management   & IEEE Explore \\
\hline
\citep{Gater+.2010}           & Model Matching          & ACM Digital \\
\hline
\citep{KimSuhh.2010}          & Design						      & ACM Digital \\
\hline
\citep{Niedermann+.2010}      & Design					        & IEEE Explore \\
\hline
\citep{SakrAwad.2010}         & Collection Management   & ACM Digital \\
\hline
\citep{TonellaDiFrancescomarino.2010} & Design				  & ACM Digital \\
\hline
\citep{Weidlich+.2010}        & Model Matching          & Matching Contest \\
\hline
\citep{Dijkman+.2011a}        & Collection Management   & Google Scholar \\
\hline
\citep{Gater+.2011}           & Model Matching          & IEEE Explore \\
\hline
\citep{Gerth+.2011}           & \multirow{2}{*}{Model Matching}  & \multirow{2}{*}{IEEE Explore} \\
\citep{Gerth.2014} 					  &                         & \\
\hline
\citep{AbbasSeba.2012}        & Collection Management   & IEEE Explore \\
\hline
\citep{Belhoul+.2012}         & Collection Management   & IEEE Explore \\
\hline
\citep{Branco+.2012}          & Model Matching  				& Matching Contest \\
\hline
\citep{Chan+.2012}						& Design							    & Google Scholar \\
\hline
\citep{Leopold+.2012}         & Model Matching          & Springer Link \\
\hline
\citep{Belhoul+.2013}         & Collection Management   & IEEE Explore \\
\hline
\citep{Dahmann+.2013}         & Business-IT Alignment   & ACM Digital \\
\hline
\citep{Klinkmuller+.2013}			& Model Matching  				& Matching Contest \\
\hline
\citep{Weidlich+.2013b}       & Model Matching  				& Springer Link \\
\hline
\citep{Weidlich+.2013}        & Model Matching  				& Matching Contest \\
\hline
\end{tabular}
\end{table}

\clearpage

\begin{table}[t]
\caption{Publications considered during the literature (part II)}
\label{tab:overviewII}
\centering
\begin{tabular}{l || c || c }
\hline
Reference & Topic & First Source\\
\hline
\hline
\citep{Baumann+.2014}         & Model Matching  				& Springer Link \\
\hline
\citep{Cayoglu+.2014}         & Model Matching  				& Matching Contest \\
\hline
\citep{Fengel.2014}           & Model Matching  				& Emeral Insight\\
\hline
\citep{KacimiTari.2014}       & Similarity Search       & IEEE Explore \\
\hline
\citep{Klinkmuller+.2014}			& Model Matching  				& Matching Contest \\
\hline
\citep{Ling+.2014}            & Model Matching  				& Springer Link \\
\hline
\citep{Baumann+.2015}				  & Model Matching  				& Springer Link \\
\hline
\citep{Belhoul+.2015}					& Collection Management   & IEEE Explore \\
\hline
\citep{LaRosa+.2015}          & Collection Management   & ACM Digital \\
\hline
\citep{SebuCiocarlie.2015}    & Collection Management   & IEEE Explore \\
\hline
\citep{Ternai+.2015}					& Design							    & Springer Link \\
\hline
\citep{Tsagkani.2015}				  & Discussion  						& Springer Link \\
\hline
\citep{Antunes+.2015}					& Process Model Matching  & Matching Contest \\
\hline
\citep{Beheshti+.2016}				& Discussion						  & Springer Link \\
\hline
\end{tabular}
\end{table}

\section{Overview of Identified Literature}
\label{sec:overview}

Next, we briefly summarize the identified publications.
To ensure that relations between these techniques are comprehensible, they are presented in historical order.

\matchertitle{Semantic Alignment of Business Processes \citep{Brockmans+.2006}}
A generic approach to identify elementary correspondences between elements of PrT nets 
is proposed in \citep{Brockmans+.2006}.
The approach identifies correspondences between transitions as well as other elements. 
First, the user determines the types of elements and properties that the approach should consider.
Next, similarity scores for all possible property pairs are calculated based on manually provided ontologies.
These property scores are then aggregated into similarity scores per element pair. 
Finally, corresponding element pairs are selected and, if necessary, another run can be triggered to refine the results.
The approach is demonstrated with regard to an example, but it is not evaluated.

\matchertitle{Matching Statecharts Specifications \citep{Nejati+.2007}}
\citet{Nejati+.2007} present a matcher that is tailored to statecharts.
Rather than matching transitions, the matcher computes correspondences between states.
There are two types of basic matchers: static matchers rely on state labels and positions; and behavioral matchers examine whether two states depend on or transition into similar states.
The evaluation resides on three model pairs for which the matcher yields recall values between $.81$ and $1.0$ and precision values between $.51$ and $.55$.

\matchertitle{Aligning Business Process Models \citep{Dijkman+.2009}} 
A configurable matching technique is introduced in \citep{Dijkman+.2009}.
It calculates the similarity of activities based on a syntactic measure where stemming can be optionally applied to unify words.
The first matcher variant proposes all activity pairs with a label similarity score higher than a threshold as a correspondence.
The second variant determines alignments by optimizing their overall similarity score. 
In particular, a graph edit distance is defined that relies on label similarity scores as well as the number of matched activities and edges.
The detection of alignments is based on the idea of minimizing the graph edit distance.
Here, a greedy search or an A-star search can be used to stepwise add activities to the alignment.
In an optional post-processing step, the matcher tries to extend elementary correspondences into complex correspondences.
The variants are assessed with regard to 17 model pairs from Dutch municipalities.
The according macro f-measures vary from $.66$ for the first variant with stemming to $.72$ for the second variant with a greedy search, but no post-processing.

\matchertitle{Complex Mapping Discovery \citep{Gater+.2010}}
Similar to \citep{Dijkman+.2009} the approach in \citep{Gater+.2010} relies on a graph edit distance and is tailored to block-structured models.
It is assumed that activities do not only possess a label, but that they are also annotated with input and output objects.
Then, the set of potential elementary correspondences comprises all activity pairs with compatible input and output objects.
Further, only pairs with a weighted similarity score that is sufficiently high are considered.
This score takes labels as well as the input and output objects into account.
Next, a set of 1:n-correspondence candidates is derived by composing elementary correspondences where the $n$ activites from one of the models need to occur in the same sequence, parallel, or exclusive block.
Finally, the alignment comprises those elementary and 1:n-correspondence candidates that together optimize a graph edit distance.
There are no evaluation results.

\matchertitle{The ICoP framework \citep{Weidlich+.2010}}
The \textit{ICoP framework} \citep{Weidlich+.2010} is a configurable and extendable matching process. 
First, \textit{searchers} which rely on similarity measures and heuristics are used to identify correspondence candidates. 
Then, \textit{boosters} are employed to narrow down the candidate set and to unify similarity scores. 
Next, \textit{selectors} construct the final set of correspondences by evaluating the similarity scores or by consulting an \textit{evaluator}.
Evaluators calculate overall scores for potential alignments and might rely on properties derived from the process models.
Different configurations of the framework that comprise specific components for each of the four types are evaluated.
In this regard, the 17 model pairs from \citep{Dijkman+.2009} are reused and three additional model pairs are introduced.
The matcher from \citep{Dijkman+.2009} is used as a baseline. 
The f-measures for all matchers differ only marginally and are located close to a value of $.6$.

\matchertitle{Summary-Based Process Model Matching \citep{Gater+.2011}}
\cite{Gater+.2011} extend their matcher from  \citep{Gater+.2010}.
Like the original matcher, the extension is tailored to block-structured models.
The matcher starts with composing parallel and alternative blocks as well as sequences into a single activities.
In this step, it also composes the labels as well as the input and output annotations.
The alignment is the sub-set of these summarized candidates that maximizes a graph edit distance.
Finally, the matcher aggregates unmatched activities into adjacent correspondences and decomposes m:n-correspondences into elementary and 1:n-correspondences.
Two variants are compared based on 1,200 model pairs and achieve an overall f-measure of about $.84$.

\matchertitle{Precise Mappings in Versioning Scenarios \citep{Gerth+.2011}}
An approach to match an original process model with two updated versions of this model is presented in \citep{Gerth+.2011,Gerth.2014}.
When a version is introduced by copying the original, the alignment is automatically established and all updates of the version result in respective alignment updates.
Yet, it is possible that the alignment between the original and the version does not contain all correspondences.
Thus, it is completed by matching activities with similar labels and edges with corresponding sources and targets.
Finally, the models are structurally decomposed into fragments whose descriptions are a combination of the labels of their activities.
By comparing these descriptions corresponding fragments are detected.
Alignments between the two versions are initially inferred from the alignments between the versions and the original.
Then, they are completed following the outlined procedure.
The effectiveness of this approach was not assessed.

\matchertitle{Matching Processes Across Abstraction Layers \citep{Branco+.2012}}
The matcher from \citep{Branco+.2012} first identifies all nodes with identical types and labels as elementary correspondences.
Then, the models are decomposed into fragment hierarchies from which complex correspondences are derived through a top-down traversal.
In this regard, for two fragments a syntactic similarity score over the union of the activity labels in the fragments is calculated. 
The technique achieves a macro f-measure of $.81$ on 110 model pairs stemming from the Bank of Northeast Brazil.
However, while the approach detects 400 of the 416 elementary correspondences, it only identifies 38 out of the 222 complex correspondences.

\matchertitle{Semantic Process Model Matching \citep{Leopold+.2012}}
Labels typically comprise an action, a business object, and additional information.
Thus, \cite{Leopold+.2012} decompose labels into these components and determine a weighted component similarity score per activity pair.
To construct an alignment based on these scores, Markov logic networks are applied.
In this regard, the construction process can be configured via different constraints that allow for including or excluding complex correspondences and for ensuring alignment consistency based on the execution semantics.
The effectiveness is assessed with regard to the publicly available university admission dataset\footnote{\url{http://www.henrikleopold.com/downloads/}, accessed: 5/4/2017.} and compared to a configuration of the ICoP framework.
Here, the ICoP framework performs slightly worse ($.318$ vs. $.294$) than the best configuration of the proposed approach.

\matchertitle{The Bag-of-Words Technique \citep{Klinkmuller+.2013}}
We introduced our purely label-based bag-of-words technique in \citep{Klinkmuller+.2013}.
For each possible activity pair it determines a similarity score by aggregating the similarity scores for the words in the activities' labels.
To account for differences in label specificity pruning can be used to eliminate words from the larger label before determining the similarity score.
Finally, all activity pairs with a similarity score higher or equal to a threshold are suggested as corresponding.
We evaluated the technique on the university admission dataset where it achieved a maximum f-measure of $.409$ which is an improvement over the values from \citep{Leopold+.2012}.
Additionally, we analyzed the false positives and negatives to identify matching challenges.
Here, we found that the correct interpretation of labels is primarily challenged by different levels of detail or abstraction, compound words as well as implicit objects and roles.

\matchertitle{The Prediction of Matching Quality \citep{Weidlich+.2013b}}
The automatic selection of matchers is discussed in \citep{Weidlich+.2013b}.
Here, the idea is to correlate the matchers effectiveness to process model and activity properties based on a set of model pairs for which the ground truth is known.
The derived prediction model can then be used to select matchers for unmatched model pairs.
The authors suggest a set of properties for models and activities based on the labels and control flow information, but do not present an evaluation.

\matchertitle{Matching Based on Positional Language Models \citep{Weidlich+.2013}}
The matcher in \citep{Weidlich+.2013} compares activities based on their labels and, if available, accompanying documenation.
First, each model is transformed into a text document by traversing a structural decomposition of the model.
Whenever an activity is reached during the traversal, a passage that consists of the label and the documentation is added to the text document.
Next, for each term in the text documents the probability of occuring in a passage is determined.
The probabilities are then used to compute similarity scores for the activity pairs and the most similar pairs are added to the alignment.
Here, different selection strategies can be applied. 
The evaluation comprises four different sets of model pairs including those from \citep{Branco+.2012} and \citep{Weidlich+.2010}.
The f-measures for different variants vary from $.18$ to $.33$ on all sets.

\matchertitle{The Process Model Matching Contest 2013 \citep{Cayoglu+.2014}}
In the first matching contest, the approaches from \citep{Dijkman+.2009,Klinkmuller+.2013,Weidlich+.2010,Weidlich+.2013b} as well as three additional approaches participated.
The \textit{Triple-S} technique adds all activity pairs with a sufficiently high similarity score to the alignment.
The similarity scores rely on a comparison of the labels as well as the number of incoming and outgoing edges.
The \textit{RefMod-Mine/NSCM} (RMM/NSCM) technique filters activities that potentially represent states or gateways through label analysis and calculates label-based similarity scores for all remaining activity pairs.
It then determines correspondences by clustering all activities in a model collection.
The \textit{RefMod-Mine/ESGM} (RMM/ESGM) technique adapts the graph edit distance approach from \citep{Dijkman+.2009}.
It additionally incorporates dictionary lookups to compare labels and completes alignments by adding activity pairs with a similarity score higher than a predefined threshold to the alignment.
The university admission \citep{Leopold+.2012} and the also publicly available birth registration dataset\footnote{\url{http://www.henrikleopold.com/downloads/}, accessed: 5/4/2017.} are used in the evaluation.
Here, all approaches yielded a low effectiveness with the highest f-measures at about $.4$.

\matchertitle{Multi-Perspective Matching \citep{Baumann+.2014}}
An extension of the matcher from \citep{Dijkman+.2009} is presented in \citep{Baumann+.2014}. 
It is limited to process models that represent sequences and differs from the original matcher in that the activity similarity measure does not only consider labels.
It also is based on the activities' position in relation to all other correspondences, the ratio of data objects shared by the activities, and the roles responsible for the execution of the activities.
The approach is demonstrated based on one exemplary model pair.

\matchertitle{Resource-Aware Process Matching \citep{Baumann+.2015}}
An extension of \citep{Baumann+.2014} is presented in \citep{Baumann+.2015}.
Here, a more fine-grain comparison of the organizational perspective, i.e., the roles that are responsible for activity execution, is considered.
The authors discuss practical limitations, but do not evaluate their matching technique.

\matchertitle{Semantic Model Alignment \citep{Fengel.2014}}
\citep{Fengel.2014} introduces an approach that solely relies on activity labels.
It comprises checks for label equality, shared words, synonyms, and negation words as well as a label based similarity.
Based on the determined similarity scores, the activity pairs are classified as an exact, a close, a loose, or a low correspondence.
Eight model pairs are used to evaluate the approach and a macro f-measure of $.89$ is reported.

\matchertitle{Adpative Label-based Matching based on Expert Feedback \citep{Klinkmuller+.2014}}
In \citep{Klinkmuller+.2014} we studied the adaptation of the matching process based on the analysis of expert feedback in terms of (in-)validated correspondences.
First, we pursued the idea of learning a classifier from expert feedback that correlates activity similarity scores to the classes of corresponding and non-corresponding activity pairs.
However, an empirical analysis based on the university admission and birth registration datasets revealed that similarity scores which are based on control flow properties do not possess a discriminative power that is high enough to separate corresponding from non-corresponding activity pairs.
Thus, we examined the idea of improving the effectiveness of our bag-of-words technique by adjusting the word similarities based on expert feedback.
We showed that this strategy leads to effectiveness improvements on both datasets of up to $53\%$.
In particular, we achieved an f-measure of $.69$ on the birth registration and of $.58$ on the university admission dataset.

\matchertitle{Fast Discovery of Complex Matches \citep{Ling+.2014}}
Another variant of the matcher by \cite{Dijkman+.2009} is introduced in \citep{Ling+.2014}.
To identify correspondences the matcher considers all activities as well as activity sets that it derives from structural decompositions of the process models as correspondence candidates.
Then, the greedy strategy from \citep{Dijkman+.2009} in combination with a modified graph edit distance is used to determine the correspondences.
The authors report an f-measure of $.73$ achieved on 20 model pairs.

\matchertitle{The Process Model Matching Contest 2015 \citep{Antunes+.2015}}
In the second matching contest, the matchers from \citep{Weidlich+.2013,Cayoglu+.2014} participated.
Additionally, nine new techniques were submitted.
The first matcher is an adapted version of the \textit{AML ontology matcher} from \citep{Faria+.2013}.
It determines correspondences based on three label similarities.
The \textit{KnoMa-Proc} matcher first joins adjacent activities to determine complex correspondence candidates.
From the union of the activities and the candidates an alignment is constructed based on a label-based confidence measure.
The \textit{Match-SSS} and the \textit{Know-Match-SSS} compare the words in the labels to compute an overall label similarity score.
Both approaches differ with regard to word similarities they rely on.
The \textit{RefMod-Mine/VM\textsuperscript{2}} (RMM/VM\textsuperscript{2}) matcher suggests equally-labeled activity pairs, pairs whose labels contain similar words in a different order, and pairs with a high label similarity score that is based on word co-occurrences in the model pair.
The \textit{RefMod-Mine/NCHM} (RMM/NCHM) is an updated version of the RMM/NSCM technique from \citep{Cayoglu+.2014} which contains an additional post-processing step to filter activity pairs with different roles.
The \textit{RefMod-Mine/NLM} (RMM/NLM) computes label similarities based on word relations in a dictionary and proposes activity pairs with a sufficiently high similarity score as correspondences.
The \textit{RefMod-Mine/SMSL} (RMM/SMSL) also investigates such word relations, but additionally optimizes the respective similarity scores based on the gold standard alignments.
Lastly, the \textit{pPalm-DS} matcher considers word occurrences in Wikipedia\footnote{\url{https://en.wikipedia.org/
}} to assess the label similarity and to suggest correspondences.
In addition to an updated version of the university admission and the original birth registration dataset, the contest evaluation comprised the asset management dataset\footnote{\url{https://ai.wu.ac.at/emisa2015/contest.php}, accessed: 5/4/2017} which we made available to the contest.
The best f-measure scores on each dataset rank in between $0.54$ and $0.68$.

\section{Analysis Results}
\label{sec:results}

Table~\ref{tab:lit:results} summarizes the results of our literature review as a matrix, where
each publication is represented by a row. 
By following guidelines for inductive category formation \citep{Mayring.2000} we classified the analysis options for incorporating control flow information into the matching process.
In total, we derived eight abstract analysis options which comprise most columns in the matrix.
They are grouped into three \textit{use cases}, and refer to a specific \textit{encoding}. 
The first use case is to \textit{compare activities} (Comp. Act.), either directly with regard to control flow properties or by incorporating structural context into the label similarity. 
Second, control flow information can be used to \textit{detect fragments} (Det. Frag.), where activity sets are derived from the model structure and considered as candidates for complex correspondences. 
Third, matchers \textit{check the consistency} (Check Consis.) to assess if control flow dependencies between activities in one model resemble those of the corresponding activities in another model. 
Per use case, there are up to three types of encoding:
Matchers might analyze path relations in the process \textit{graph} (G), 
properties of nested \textit{fragment hierarchies} (H), or \textit{execution semantics} (ES).
The cells in the columns of Table~\ref{tab:lit:results} capture which abstract analysis options occur in which publications (occurring/not as ``+''/``-''). 
Note that our identification did not require matchers to rely on control flow information, hence four of the papers contain none.

\begin{table}[t]%
\centering
\caption{Overview of options to integrate control flow information into the matching process and the empirical evidence}
\label{tab:lit:results}
\begin{tabular}{l | C{.6cm} C{.6cm} C{.6cm} | C{.6cm} C{.6cm} | C{.6cm} C{.6cm} C{.6cm} || l}
\toprule
                & \multicolumn{3}{c|}{\textit{Comp. Act.}}& \multicolumn{2}{c|}{\textit{Det. Frag.}}& \multicolumn{3}{c||}{\textit{Check Consis.}} & \textit{Paper}\\
\textit{Source} & \textit{G} & \textit{H} & \textit{ES} & \textit{G} & \textit{H} & \textit{G} & \textit{H} & \textit{ES} & \textit{Class}\\
\midrule
\citep{Brockmans+.2006}       & - & - & -    & - & -    & - & - & -     & Illustrated\\ 
\citep{Nejati+.2007}         & + & - & -    & - & -    & - & - & +     & Compared\\ 
\citep{Dijkman+.2009}        & - & - & -    & + & -    & + & - & -     & Compared\\ 
\citep{Gater+.2010}          & - & - & -    & - & -    & + & - & -     & Proposed\\ 
\citep{Weidlich+.2010}       & + & + & -    & + & +    & + & + & -     & Compared\\ 
\citep{Gater+.2011}          & - & - & -    & + & +    & + & - & -     & Compared\\ 
\citep{Gerth+.2011,Gerth.2014}          & - & - & -    & + & +    & - & - & -     & Proposed\\ 
\citep{Branco+.2012}         & - & - & -    & - & +    & - & - & -     & Compared\\ 
\citep{Leopold+.2012}        & - & - & -    & - & -    & - & - & +     & Compared\\ 
\citep{Cayoglu+.2014}        & + & + & -    & + & +    & + & + & + 		& Evaluated\\ 
\citep{Klinkmuller+.2013}    & - & - & -    & - & -    & - & - & - 		& Compared\\ 
\citep{Weidlich+.2013b}      & - & - & -    & - & -    & - & - & - 		& Proposed\\ 
\citep{Weidlich+.2013}       & + & + & -    & - & -    & - & - & - 		& Compared\\ 
\citep{Baumann+.2014}        & + & - & -    & + & -    & + & - & - 		& Illustrated\\ 
\citep{Baumann+.2015}        & - & - & -    & - & -    & + & - & - 		& Proposed\\ 
\citep{Fengel.2014}          & - & - & -    & - & -    & - & - & - 		& Evaluated\\ 
\citep{Klinkmuller+.2014}    & + & + & +    & - & -    & - & - & - 		& Analyzed\\ 
\citep{Ling+.2014}           & + & - & -    & - & +    & + & - & - 		& Evaluated\\ 
\citep{Antunes+.2015}        & + & - & -    & - & -    & + & - & - 		& Evaluated\\ 
\bottomrule
\end{tabular}
\end{table}

The last column characterizes the empirical evidence that is given towards the proposed matchers and thus towards the successful use of control flow information. 
As we are interested in the use of control flow information, we discard the four publications that solely exploit labels and focus on the 15 publications where control flow information is considered in the following. 
Three papers only \textit{propose} matchers but provide no empirical evidence, and another paper uses one synthetic example to \textit{illustrate} how the matcher is supposed to work.
Among the remaining 11 publications, there are three papers \citep{Cayoglu+.2014,Antunes+.2015,Ling+.2014} 
that \textit{evaluate} matchers as black boxes.
Such an evaluation assesses the effectiveness of entire matchers, but does not study the influence that the matchers' components have on the effectiveness.
The matcher in \citep{Ling+.2014} comprises components that compute label similarity scores, investigate the graph neighborhood, detect fragments, and check the consistency.
Clearly, the reported overall effectiveness 
allows no insights into the contribution of each component. 
Similarly, the contests \citep{Cayoglu+.2014,Antunes+.2015} compare the effectiveness of various matchers, but they do not 
provide any insights on how the results of individual matchers are influenced by their components. 
Another seven papers \textit{compare} the effectiveness of different matcher variants.
However, as e.g., discussed in \citep{Salzberg.1997,Demsar.2006}, such results need to be interpreted with care and typically have a limited validity. 
That is because without further statistical analyses differences might have been observed simply by chance -- especially as the reported difference are rather small, e.g., the f-measures in \citep{Dijkman+.2009} differ by $\approx.06$ and in \citep{Weidlich+.2010} by $\approx.05$.
Moreover, the results of all variants are typically dependent on a basic variant.
This entails the risk that the relative performance of the variants and thus the contribution of the propositions changes, if the basic variant is modified.
For example, consistency checks improve the effectiveness in \citep{Leopold+.2012}, whereas they reduce it in \citep{Weidlich+.2010}.
Finally, our prior work \citep{Klinkmuller+.2014} is the only paper that explicitly analyzes the validity of control flow propositions -- but only for the first use case.

Overall, the review revealed that despite the pervasive use of control flow information in the literature, the evidence towards the positive impact of exploiting such information is limited.
Thus, to comprehensively assess the validity of the analysis options, additional analysis is warranted.
\section{Summary}
\label{sec:conclusion}

In this report, we examined publications that introduced process model matching techniques.
In particular, we focused on understanding, if and how control flow information contributes to the identification of corresponding activities between process models.
Our analysis revealed that there are three dominant use cases for the analysis of control flow information and that three different types of encodings are used.
However, while many matchers rely on control flow information, the evidence given towards the successful contribution of this information to the matching process is limited.
That is because the majority of the publications focused the effectiveness evaluation, but did not carefully study how the analysis of control flow information contributes towards the effectiveness.
There is one notable exception.
In our own work \citep{Klinkmuller+.2014} we studied the comparison of activities with regard to control flow properties and revealed that such a comparison is not suited for informing matching techniques.
Overall, this result shows that more research is needed in order to understand the effects of relying on control flow information.

\bibliography{references}

\end{document}